# Aging cellular networks: chaperones as major participants

C. Sőti, and P. Csermely[1]

Department of Medical Chemistry, Semmelweis University,
P. O. Box 260, H-1444 Budapest 8, Hungary

**Abstract**
We increasingly rely on the network approach to understand the complexity of cellular functions. Chaperones (heat shock proteins) are key "networkers", which have among their functions to sequester and repair damaged protein. In order to link the network approach and chaperones with the aging process, we first summarize the properties of aging networks suggesting a "weak link theory of aging". This theory suggests that age-related random damage primarily affects the overwhelming majority of the low affinity, transient interactions (weak links) in cellular networks leading to increased noise, destabilization and diversity. These processes may be further amplified by age-specific network remodelling and by the sequestration of weakly linked cellular proteins to protein aggregates of aging cells. Chaperones are weakly linked hubs [i.e., network elements with a large number of connections] and inter-modular bridge elements of protein-protein interaction, signalling and mitochondrial networks. As aging proceeds, the increased overload of damaged proteins is an especially important element contributing to cellular disintegration and destabilization. Additionally, chaperone overload may contribute to the increase of "noise" in aging cells, which leads to an increased stochastic resonance resulting in a deficient discrimination between signals and noise. Chaperone- and other multi-target therapies, which restore the missing weak links in aging cellular networks, may emerge as important anti-aging interventions.

**1. Introduction: aging cellular networks**

The tremendous increase of our knowledge on the individual pathways and interactions in cells and the integration of this knowledge to a number of annotated databases allow and require a holistic approach in order to recognize the emerging properties of this complexity. This approach is conveniently served by the network description of the cellular organization.

Cellular networks can be divided into two major types each with three divisions. The first includes three networks (protein-protein interaction networks, cytoskeletal networks and membrane-organelle networks) which summarize our knowledge of hierarchic cellular architecture (Fig. 1). In the protein-protein interaction network the elements of the network are proteins and the links between them are permanent or transient bonds (von Mering et al., 2002; Rual et al., 2005; Stelzl et al., 2005; Gavin et al., 2006). It is important to note that in this network the interactions are averaged, i.e., an interaction between protein A and protein B denotes the probability of the interactions of the actual copies of these two proteins. In the cytoskeletal network, individual cytoskeletal filaments, like actin, tubulin filaments, or their junctions as the elements of the network, and the bonds between them are the links. In the membrane-organelle network various membrane vesicles and cellular organelles are the elements and usually protein complexes link them together. Each of the elements of this latter network contains large protein-protein interaction networks. The second network also includes three networks (the gene transcription network, the signalling network and the metabolic network) which provide information on the complex dynamism of the cell (Fig. 1). Gene transcription networks are correlation networks, where the elements are genes and many of their interactions are mostly established from the correlations in their expression patterns. (This information is extended by our biochemical knowledge on the regulation of gene transcription by various transcriptional factors; Blais and Dynlacht, 2005). Metabolic and signalling networks describe the conformational changes of cellular proteins. Metabolic networks summarize the most frequent conformational changes we call enzyme activity, while signalling networks provide a framework for the description of infrequent conformational changes we call signal transduction (Borodina and Nielsen, 2005; White and Anderson, 2005; Irish et al., 2006).

---

[1]To whom all correspondence should be addressed: csermely@puskin.sote.hu (www.weaklinks.sote.hu)

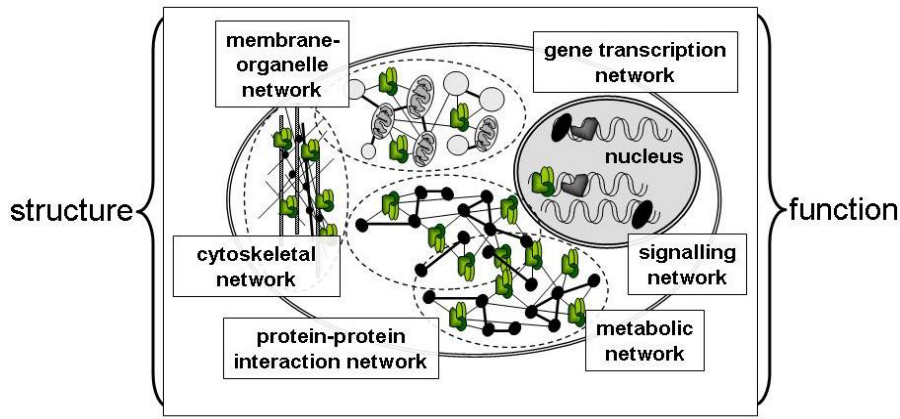

**Figure 1.** Cellular networks. The protein-protein interaction network, the cytoskeletal network and the membranous, organelle networks provide a hierarchical scaffold of the cell. The other three networks, like the transcriptional, metabolic, or signaling networks are functionally defined. Elements of the transcriptional networks are genes and the connecting links are functional interactions between them. Many of these interactions are derived from gene co-expression patterns. The metabolic and signaling networks summarize the conformational changes of cellular proteins. The metabolic networks describe the most frequent changes of the enzyme, while the links of signaling networks denote the more infrequent conformational changes of signal transduction. All these networks highly overlap with each other, and some of them contain modules of other networks.

Cellular networks have four major tasks. (1) The first task is the confined dissipation of the noise. (2) The second task is the fast, targeted and distortion-free transmission of signals to their distant destinations. (3) The third task is the discrimination between signals and noise via the continuous remodelling of these networks during the evolutionary learning process of the cell. (4) The fourth task is the protection against the continuous random damage of free radicals and other harmful effects during stress and aging (Csermely, 2006).

Aging is accompanied by a general increase in noise parallel with a decrease in complexity (Hayflick, 2000; Goldberger et al., 2002; Herndon et al., 2003). The seminal paper of Himmelstein et al. (1990) summarizes the erosion of homeostatic capacity during aging. The loss of complexity (system integrity) will lead to increased noise and vice versa, increased noise resulting in decreased repair and remodelling processes which results in a loss of complexity. These two phenomena are different sides of the same coin and may form a vicious circle aggravating the status of the aging body. Noise can be useful in the recognition of signals, which are only slightly below a signal transduction threshold. This phenomenon is called either stochastic resonance or stochastic focusing, depending on whether the noise is extrinsic or intrinsic, respectively. If noise is low, a low level signal may not pass the signal transduction threshold. On the contrary, if the noise is high, the same low level signal will pass the same threshold a higher fraction of the time (Fig. 2; Paulsson et al., 2000).

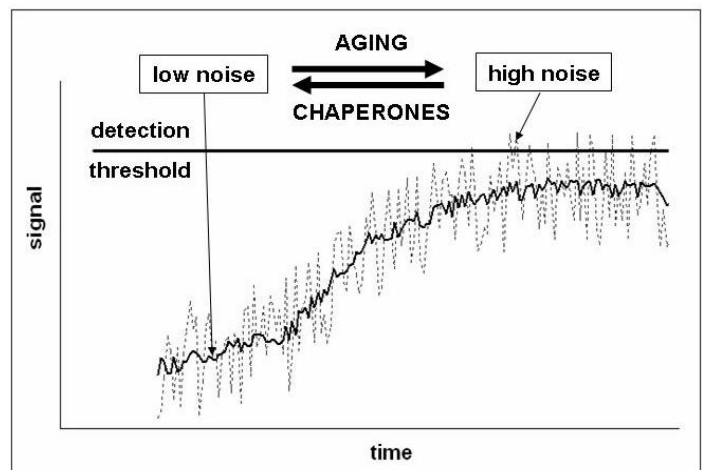

**Figure 2.** Aging and chaperones as opposing forces in the regulation of stochastic resonance. Stochastic resonance is the phenomenon, where intrinsic or extrinsic noise helps a low level signal to surpass a signal transduction threshold. Aging induces a higher noise level, which helps the recognition of sub-threshold signals. On the contrary, molecular chaperones may help to decrease the noise of cellular processes, which may prevent low-level signals to act.

In the network theory, aging is usually described as chronic random damage leading to as rapid deterioration of emergent network properties (Dorogovtsev and Mendes, 2000; Zhu et al., 2003; Agoston et al., 2005). Chan et al. (2004) showed that a rapidly aging network makes the older nodes rather isolated and only a limited local spread of information instead of global coupling. This property of aging networks may contribute to the loss of integration in aged organisms. Hubs (network elements with a large number of connections) seem to be especially important in the aging process, since proteins associated with aging are preferentially hubs (Promislow, 2004), and the function of an aged network can be significantly improved by a limited intervention via its hubs (Ferrarini et al., 2005).

Previously Promislow (2004) reported that most of the age-induced random damage affected links between weakly connected proteins due to large part that these links are the most common. However, an important question arises here. What would result, if a large number of weak links are deleted from a network? Weak links stabilize a vast variety of networks including all those listed at the beginning of this review. It is a general phenomenon that the loss of weak links does not affect the major responses of these networks. However, numerous examples prove that deletion of weak links leads to an unbalanced system, with a larger noise and instability (Csermely, 2004; 2006). These major properties of aging networks, have been mentioned previously. Thus, a "weak link theory of aging" can be proposed, where the preferential loss of low affinity, transient, low probability interactions of cellular networks emerges as an important mechanism explaining the increase in noise parallel with a decrease in complexity in aging organisms. An additional mechanism for the "weak link theory of aging" may be that the aging cell gradually loses segments of its original function due to the permanent random damage and insufficient repair. Consequently, the shrinking resources of the aging cell withdraw the pathways to the most important, vital routes (the so-called "skeleton" of the network; see Song et al., 2005 or for a related remodelling of metabolic networks during reductive evolution: Pal et al., 2006) and preferentially lose the secondary, weak pathways, which were back-ups, or gave an extra stability. This age-related cellular remodelling again induces a preferential loss of weak links in cellular networks leading to the instability and noise mentioned above. Aging often results in an increase in protein aggregation, like in the prevalent neurodegenerative diseases, such as in Huntington's disease, Alzheimer's disease or Parkinson's disease (Soti and Csermely, 2002). Aggregating proteins may sequester a number of cellular proteins by forcing their co-aggregation in a rather unspecific, capture-like process (Donaldson et al., 2003). It is possible, that predominantly those proteins with weak links (low affinity interactions) can be captured along with their original partners. Based on this logic, protein aggregation may also preferentially decrease the amount of weak links in an aging cell, leading to the very same decrease in stability as the damage-induced loss of weak links, or the remodelling-related loss of weak links induced by the shrinking resources of the aging cell. The proof of this "weak link theory of aging" requires further studies on the position of aging genes in cellular networks, on the distribution of the strength of their contacts, and on the effects of permanent random damage as well as network rearrangements during this process. However, even in its current, preliminary form the theory shows the strength of the connection of network topology and dynamics with the emergent network properties to explain the complex aging phenotype.

## 2. Molecular chaperones of aging cells

Chaperones either assist in folding of newly synthesized or damaged proteins in an ATP-dependent, active process, or work in an ATP-independent, passive mode sequestering damaged proteins for future refolding or proteolysis (Young et al., 2004). Environmental stress leads to proteotoxic damage. Damaged, misfolded proteins bind to chaperones, and liberate the heat shock factor (HSF) from its chaperone complexes. HSF is activated and transcription of chaperone genes takes place (Morimoto, 2002). Most chaperones, therefore, are also called stress or heat shock proteins. As a function on the level of the whole organism, chaperones act as a buffer to conceal the phenotype of the genetic changes in a large variety of organisms (Rutherford and Lindquist, 1998; Fares et al., 2002; Queitsch et al., 2002; Cowen and Lindquist, 2005).

In aging organisms protein damage becomes prevalent: e.g., in an eighty year old human half of all proteins are estimated to be oxidized (Stadtman and Berlett, 1998). Damaged proteins compete with HSF in binding to the Hsp90-based cytosolic chaperone complex, which may contribute to the generally observed constitutively elevated chaperone levels in aged organisms (Zou et al., 1998; Soti and Csermely, 2003). On the contrary, the majority of the reports showed that stress-induced synthesis of chaperones is impaired in aged animals. While HSF activation does not change, DNA binding activity may be reduced during aging (Heydari et al., 2000). Initial studies indicate that chaperone function also becomes compromised in aged cells (Cherian and Abraham, 1995; Nardai et al., 2002). Two independent reasons emerge as possible reasons of age-induced chaperone dysfunction. Chaperones may become preferentially damaged in aging (Macario and Conway de Macario, 2005) acting as "suicide proteins" and becoming "sick chaperones". As an example of this, bacterial homologues of Hsp60 and Hsp70 are preferentially oxidized in growth-arrested E. coli (Dukan and Nyström, 1999). Another possible reason of decreased chaperone function is chaperone overload. Chaperone overload occurs, where the need for chaperones may greatly exceed the available chaperone capacity, which probably happens in most aged organisms. This may cause defects in signal transduction, protein transport, immune recognition, cellular organization as well as the appearance of previously buffered, hidden mutations in the phenotype of the cell (Csermely, 2001; Nardai et al., 2006). Chaperone overload may significantly decrease the robustness of cellular networks, as well as shift their function towards a more stochastic behavior. As a result of this, aging cells become more disorganized, their adaptation is impaired.

### 3. Chaperones as key elements of aging networks

Chaperones form extensive complexes and have a large number of co-chaperones to regulate their activity, binding properties and function (Young et al., 2004; Blatch, 2006). These chaperone complexes regulate local protein networks, such as the mitochondrial protein transport apparatus as well as the assembly and substrate specificity of the major cytoplasmic proteolytic system, the proteasome (Imai et al., 2003; Young et al., 2003). Chaperones may be important elements to promote the cross-talk between various signaling pathways. As an example of this, the Hsp90 chaperone complex promotes the maturation of over a hundred kinase substrates (Zhao et al., 2005; Nardai et al., 2006; Rutherford et al., 2006).

Chaperones have an unusually large proportion of hubs among their neighbors in the yeast protein-protein interaction network (Csermely et al., 2006), which gives them a central position in the network topology. Moreover, key chaperones connect functional modules of the yeast protein-protein network, which further substantiates their central position of cellular processes (Fig. 3). Interestingly, the current examples of inter-modular chaperones (such as SSA4, SSE2 or Hsp82 on Fig. 2) are not the major chaperones, which protect against the proteotoxic effects of stress, but are acting as a "second line of defense" becoming mobilized only in cells, where the "primary" chaperones (such as most of the SSA class of Hsp70 chaperones) have been inhibited (Werner-Washburne et al., 1983; Mukai et al., 1993; Yam et al., 2005). Importantly, some of these chaperones, like Hsp82 have been involved in the chaperone-mediated buffering described above (Cowen and Lindquist, 2005). This raises the hypothesis that chaperones may be divided to two classes, where the "primary defense" acts as a conventional chaperone, and protects the cells against damage, while the "second line of defense" reconfigures cellular networks. The examination of this hypothesis awaits further research.

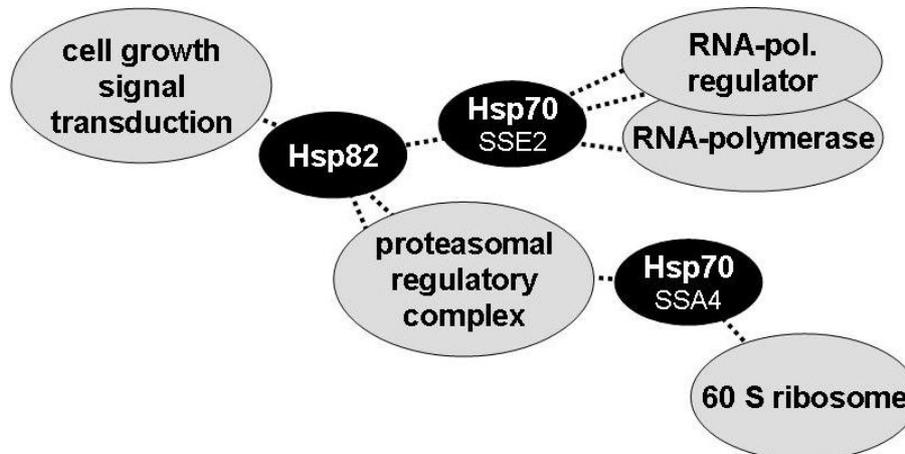

**Figure 3.** Molecular chaperones as inter-modular, central elements of the yeast protein-protein interaction network. Chaperone interactions were assessed using the annotated database of von Mering et al. (2002) and protein modules were those established by Valente (2005).

Chaperones are typical weak linkers, providing low affinity, low probability contacts with other proteins. Weak links are known to help system stability in a large variety of networks from macromolecules to social networks and ecosystems (Csermely, 2004; 2006), which may be a general network-level phenomenon explaining many of the chaperone-mediated buffering effects. By decreasing the "noisiness" of cellular signals, well functioning molecular chaperones might also decrease the sensitivity of signal transduction thresholds. As a consequence, well-functioning chaperones make low intensity signals undetectable, which may also contribute to their "buffering" effects (Rutherford et al., 2006). This effect might be reversed in aging organisms, where chaperones became damaged and overloaded, which relieves their noise buffering. The increasing noise level induces an improved recognition of low-intensity signals by an increased stochastic resonance (Fig. 2). This results in a "second-order" increase in the noise level, since those signals, which were recognized as noise and were filtered out by the intensity threshold, now may go above the same threshold.

Extending the inter-modular position of chaperones introduced before, a two-hybrid study using the voltage-dependent anion-selective channel (VDAC) identified mtHsp70/Grp75, a major mitochondrial matrix chaperone as a partner of VDAC. Moreover, Grp75 provided a contact between the IP3 receptor and VDAC and reduced the cation selectivity of VDAC, thus protected mitochondria from Ca2+ overload and permeability transition. Interestingly, Grp78 which normally resides in the ER translocated to the mitochondria and became part of the VDAC-complex after heat shock (Fig. 4; Schwarzer et al., 2002; He and Lemasters, 2003; Szabadkai and Rizzuto, 2006). Based on these and other findings Szabadkai and Rizzuto (2006) proposed that chaperones might both maintain the communication between the endoplasmic reticulum (ER) and mitochondria as well as protect these organelles from excessive death signals.

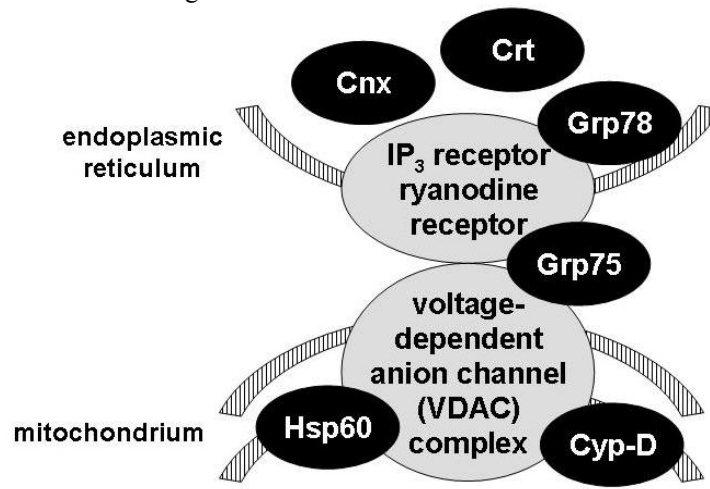

**Figure 4.** Molecular chaperones as potential regulators of the coupling of mitochondrial networks and the endoplasmic reticulum. The figure has been drawn using the data of Schwarzer et al. (2002), He and Lemasters (2003), as well as Szabadkai and Rizzuto (2006). Cnx, calnexin; Crt, calreticulin; Grp75 and Grp78, 75 kDa and 78 kDa glucose-regulated proteins, respectively; Hsp60, mitochondrial 60 kDa heat shock protein; Cyp-D, cyclophilin-D.

Generalizing the above scenario, de-coupling of network elements and modules is a widely used method to stop the propagation of damage (Barabasi and Oltvai, 2004; Csermely, 2006). During stress, chaperones become increasingly occupied by damaged proteins. This, together with the stress-induced translocation of chaperones to the nucleus (Nollen et al., 2001), might lead to an "automatic" de-coupling of all chaperone-mediated networks including protein-protein, signaling, transcriptional regulatory as well as membranous, organellar networks providing an additional safety measure for the cell (Soti et al., 2005a).

As a summary, aging-induced damage and overload of chaperones might lead to the following network damage:
- de-coupling of important modules of protein-protein interaction networks and signaling networks;
- de-coupling of mitochondrial networks, ER-mitochondrial interactions, and a general des-organization of membrane traffic and networking of the aging cell.

These changes attenuate the propagation of the signals, and consequently make the signals noisier. Chaperone damage and overload in the aging cell induces a decreased stress-resistance, and the disintegration, stochastic responses, noise and diversity of the aging cell grow in agreement with the general characterization of the aged organisms above.

## 4. Summary and perspectives

The "weak link theory of aging" presented in this paper has three elements, which all lead to an increased noise, destabilization and diversity of the aging cells: (1) the age-related random damage primarily affects the overwhelming majority of the low affinity, transient interactions (weak links) in the cellular networks; (2) an age-specific network remodelling is induced by the shrinking resources, which primarily shuts down the secondary, weak pathways, which were back-ups, or provided extra stability; (3) the sequestration of weakly linked cellular proteins to protein aggregates preferentially decreases the amount of weak links in an aging cell.

Chaperones are weakly linked hubs and inter-modular bridge elements of protein-protein interaction, signalling and mitochondrial networks. These key elements are increasingly overloaded by damaged proteins as aging proceeds, which makes them an especially important element of cellular disintegration and destabilization. Chaperone overload may also contribute to the increase of noise in aging cells, which leads to an increased stochastic resonance resulting in a deficient discrimination between signals and noise. There are several important aspects for future investigation of this field:
- the extent of age-induced damage of strong and weak links has yet to be established;
- there is little detailed information on the remodeling of aging cellular networks;
- the exact measurement of increased cellular noise levels during the aging process is missing;

- there is little information on the extent of stochastic resonance in aging;
- detailed proteomic analysis of protein aggregates has yet to be performed;
- finally, our knowledge on the exact role of chaperones in cellular networks is in its infancy.

The restoration of the missing weak links in the aging cells may be a key issue of successful anti-aging therapies. Therapies, which help repair the target-specific induction of molecular chaperones where needed (Westerheide and Morimoto, 2005; Soti et al., 2005b), and other multi-target therapies modifying a large number of low-affinity interactions (Csermely et al., 2005) all increase the number of weak links by either making new low intensity interactions, or by converting former high intensity interactions to low intensity ones. We are just at the beginning of system-level anti-aging interventions with the goal of improving the context of cellular networks as opposed to the restoration of a few major pathways.

**Acknowledgements**

Work in the authors' laboratory was supported by research grants from the EU 6th Framework program (FP6506850, FP6-016003), Hungarian Science Foundation (OTKA-F47281) and from the Hungarian National Research Initiative (1A/056/2004 and KKK-0015/3.0). C.S. is a Bolyai research Scholar of the Hungarian Academy of Sciences.